# Electric field control of disorder-tunable superconductivity and the emergence of quantum metal at an oxide interface

Zheng Chen[1†], Yuan Liu[1†], Hui Zhang[2†], Zhongran Liu[3], He Tian[3], Yanqiu Sun[1], Meng Zhang[1], Yi Zhou[4,5,6]*, Jirong Sun[4,5]*, and Yanwu Xie[1,7]*

[1]Interdisciplinary Center for Quantum Information, Zhejiang Province Key Laboratory of Quantum Technology and Device, Department of Physics, Zhejiang University, Hangzhou 310027, China.

[2]Fert Beijing Institute, School of Microelectronics, BDBC, Beihang University, Beijing, 100191, China.

[3]Center of Electron Microscope, State Key Laboratory of Silicon Materials, School of Materials Science and Engineering, Zhejiang University, Hangzhou, 310027, China.

[4]Beijing National Laboratory for Condensed Matter Physics & Institute of Physics,

Chinese Academy of Sciences, Beijing 100190, China.

[5]Songshan Lake Materials Laboratory, Dongguan, Guangdong 523808, China.

[6]Kavli Institute for Theoretical Sciences and CAS Center for Excellence in Topological Quantum Computation, University of Chinese Academy of Sciences, Beijing 100190, China.

[7]Collaborative Innovation Center of Advanced Microstructures, Nanjing University, Nanjing 210093, China.

*Correspondence to: yizhou@aphy.iphy.ac.cn (Y.Z.); jrsun@iphy.ac.cn (J.S.); ywxie@zju.edu.cn (Y.X.);

†These authors contributed equally to this work.

**We report on an extraordinary field effect of the superconducting $LaAlO_3/KTaO_3$(111) interface with $T_c$ ~2 K. By applying a gate voltage ($V_G$) across $KTaO_3$, the interface can be continuously tuned from superconducting into insulating states, yielding a dome-shaped $T_c$-$V_G$ dependence. The electric gating has only a minor effect on carrier density as evidenced in the Hall-effect measurement, while it changes spatial profile of the carriers in the interface, hence the carrier's disorder level significantly. As temperature is decreased, the resistance saturates at lowest temperature in both superconducting and insulating sides, despite an initial dramatic dropping or increasing, which suggests an emergence of quantum metallic state associated with failed superconductor and/or fragile insulator. A $V_G$-modulation of the magnetic-field-driven superconductor to insulator quantum phase transition reveals a non-universal criticality.**

**Introduction**

Controlling superconductivity with an electric field is intriguing for both fundamental researches and potential applications (*1-11*). In analogy to semiconducting field-effect transistors (*9*), the two-dimensional (2D) carrier density $n_{2D}$ of a superconductor can be tuned by applying an external gating voltage $V_G$. The attainable tuning ability can be estimated as $\delta n_{2D} = \varepsilon_0 \varepsilon_r V_G / t$, where $\varepsilon_0$ is the vacuum permittivity, $\varepsilon_r$ and $t$ are the dielectric constant and the thickness of a dielectric material respectively. To achieve a prominent tuning of superconductivity, particularly a transition from superconducting to insulating states, a $\delta n_{2D}$ that is comparable with $n_{2D}$ is generally needed. Unfortunately, the $n_{2D}$ of most superconductors is very high and far beyond the capability of a typical electric gating (~$10^{14}$ cm$^{-2}$ or lower) (*9*). Therefore, it has been a long-standing challenge to control superconductivity electrostatically.

With the recent advances of material fabrication and tuning technology, so far prominent tunings of superconductivity have been experimentally realized for a few systems, including oxide interfaces (*1, 6, 7*), electric-double-layer interfaces (*2–4, 8, 10, 11*), and magic-angle graphene superlattice (*5*). These tunings were achieved by pushing the superconducting layer to an ultrathin limit to reduce $n_{2D}$, or by developing new technologies such as electric-double-layer gating to improve tuning ability (*3, 11*), or more often, by both (*2, 12*). The tuning parameter in all these systems is $n_{2D}$.

KTaO$_3$(KTO) shares many common properties with SrTiO$_3$ (*13*). In comparison with the well-known LaAlO$_3$(LAO)/SrTiO$_3$ interface (*14*), a 2D electron gas was observed at LAO/KTO(001) (*15*) and LaTiO$_3$/KTO(001) (*16*) interfaces, but no superconductivity was reported. Superconductivity was observed at the electric-double-layer-gated KTO(001) surface, however, the transition temperature $T_c$ is low (~0.05 K) (*4*). Very recently, it was found that the EuO/KTO(111) and LAO/KTO(111) interfaces can be superconducting with a $T_c$ up to ~2 K (*17*).

In this work, we demonstrate that the newly discovered superconducting LAO/KTO(111) interface is a remarkable system that can be tuned by applying a $V_G$, in which the key tuning parameter is the disorder level (or the mobility of charge carriers) rather than $n_{2D}$. Versatile quantum phenomena are revealed consequently.

**Methods**

We grew LAO films by pulsed laser deposition on 0.5-mm-thick single-crystalline KTO(111) substrates (*18*). The surface of these films is atomic-level flat (Fig. S1). The microstructure of the interface was assessed by scanning transmission electron microscopy (STEM), which shows that LAO is homogeneous and amorphous (Figs. 1A & 1B). The lack of epitaxial growth can be ascribed to the large lattice mismatch (*15*): the cubic lattice constants are 0.379 and 0.399 nm for LAO and KTO, respectively. The LAO films themselves are highly insulating; the conductance is in the KTO layer near the interface. A sketch of the field-effect device is shown in Fig. 1C. The gating voltage $V_G$ is applied between the conducting LAO/KTO layer and a silver paste electrode at the back of KTO. The active area of the device was patterned into a Hall bar configuration (Fig. 1D). All transport measurements in this study were performed on a same device.

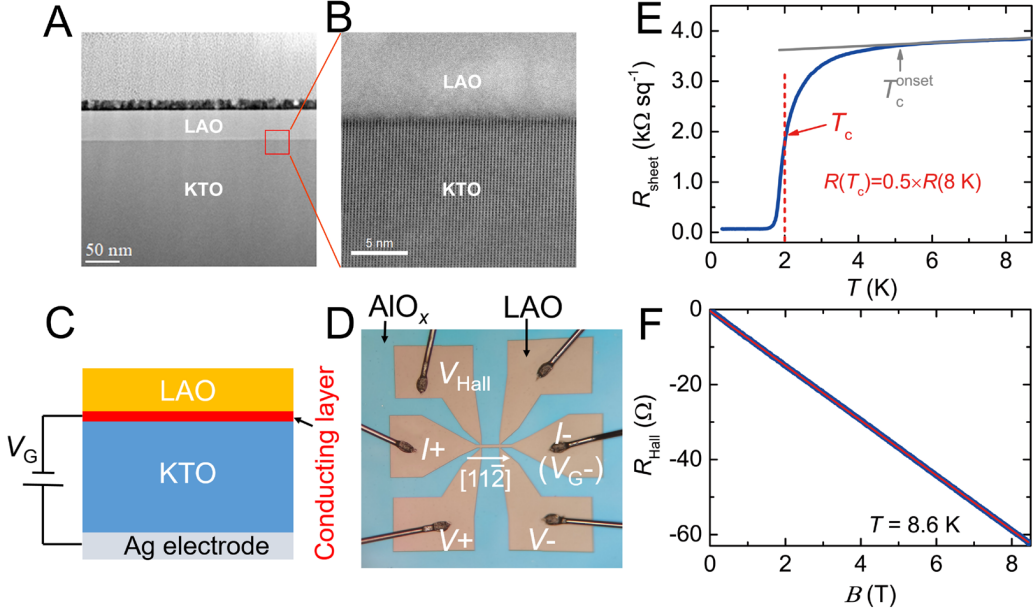

**Fig. 1. Characterizations of LAO/KTO(111) device.** (**A, B**) STEM images of a 40-nm LAO/KTO sample show that LAO film is homogeneous and amorphous. (**C**) Cross-section view of device structure. (**D**) Photograph of a typical device with bonded Al contacting wires. The device was fabricated in a six-probe Hall bar configuration. The conducting region is limited to the LAO/KTO interface (Hall bar area). The outside insulating region was obtained by coating amorphous $AlO_x$ as hard mask. The central Hall bar bridge is 20 μm in width and 100 μm in length (between $V^+$ and $V^-$). The current drain ($I^-$) and the gating contact to the interface ($V_G$-) share a same probe. The electrical current flows along $[11\bar{2}]$ crystal axis. (**E**) Dependence of $R_{sheet}$ on temperature and (**F**) dependence of $R_{Hall}$ on magnetic field for the 20-nm LAO/KTO device at $V_G = 0$.

## 2D superconductivity

The occurrence of superconductivity with a mid-point $T_c \sim 2$ K can be clearly seen in the temperature-dependent sheet resistance $R_{sheet}(T)$ (Fig. 1E). The thickness of the superconducting layer is estimated to be ~4 nm, while the estimated coherence length is ~18.8 nm (Figs. S3&S10). The comparison of these two length scales suggests a 2D superconductor. The 2D nature of the superconductivity can be further justified by the nonlinear current-voltage response ($V \sim I^3$ dependence indicates a typical BKT transition as shown in Fig. S4), the large anisotropy of upper critical field $H_{c2}$ (Fig. S3), and the magnetic-field-angle dependence of $R_{sheet}(T)$ (Fig. S5).

## $V_G$-modulation on transport properties

Normal-state Hall resistance measured at 8.6 K (Fig. 1F) demonstrates that the charge carriers are electrons rather than holes, and the carrier density $n_{2D}$ is ~8.5×10$^{13}$ cm$^{-2}$. This value is much higher than that of a typical LAO/SrTiO$_3$ interface (*1, 19–22*), and is not expected to be significantly tuned by applying $V_G$. As an estimation, adopting a low-temperature $\varepsilon_r$ of 5000 for KTO (*23*), a 300-V variation in $V_G$ can only induce a $\delta n_{2D}$ of ~1.7×10$^{13}$ cm$^{-2}$ (~20% of the total $n_{2D}$). Given the fact that $\varepsilon_r$ in the interface region can be severely degraded by growth-induced defects and $V_G$-caused electric field, the real attainable density $\delta n_{2D}$ should be even smaller, and thus insufficient to produce any prominent tuning. However, in reverse to this expectation, the transport properties of the device have been strongly tuned. With sweeping $V_G$ from 150 V to -200 V, the device transits continuously from a superconductor to an insulator, and the normal-state resistance varies for more than two orders of magnitude (Fig. 2A). The $V_G$-tuned superconductor-insulator transition can also be well seen in the evolution of current-voltage behaviors (Fig. 3). The mid-point $T_c$ derived from the $R_{sheet}(T)$s (Figs. 2A, S6 & S7) exhibits an interesting dome-shaped dependence on $V_G$ (Fig. 2B). This dependence apparently resembles the dome-shaped superconducting phase diagram observed in many other systems(*1, 4, 11*), but the underneath mechanism is different here, because $n_{2D}$ in the present device is not significantly changed by $V_G$, as will be shown below.

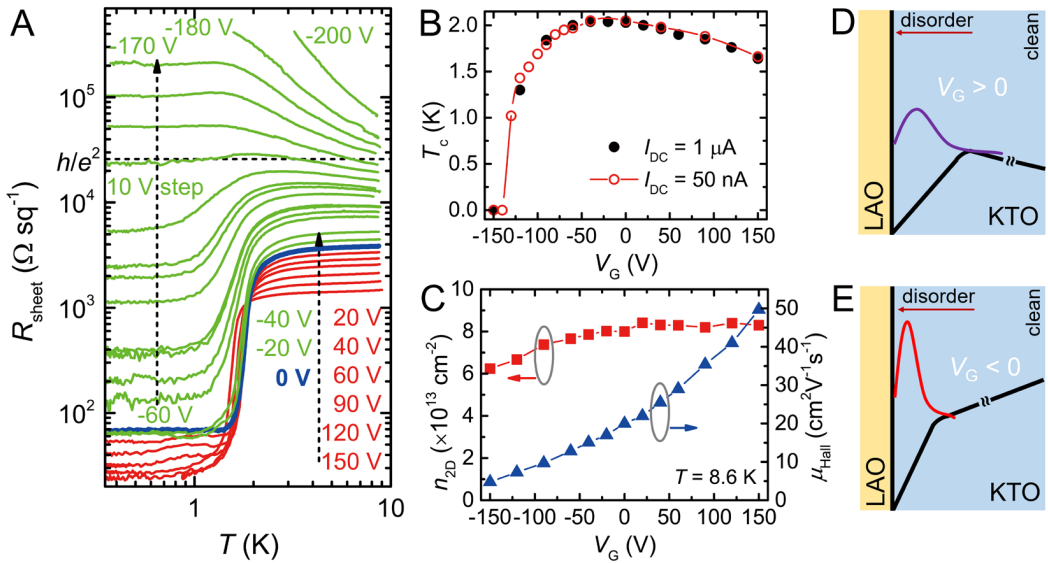

**Fig. 2. Electric Field control of transport properties.** (**A**) Temperature-dependent $R_{sheet}(T)$, (**B**) mid-point $T_c$, and (**C**) carrier density $n_{2D}$ and mobility $\mu_{Hall}$ at different $V_G$s. The dashed line in (A) indicates the position of the quantum resistance $h/e^2$. Two sets of data measured using different DC currents are shown in (B). The black filled circles represent the data measured in the same run with the Hall effect measurements. (**D, E**) Schematic diagrams of the operations for $V_G > 0$ and $V_G < 0$, respectively. The purple and red lines represent the electron envelope wave function perpendicular to the interface, matching the potential well (thick solid lines) at the corresponding $V_G$. Higher disorder strength is expected in the region closer to the interface.

**Electrostatic tuning of disorder level**

To explore the essence of these tunings, we carried out normal-state Hall effect measurements for different $V_G$s. To avoid any hysteresis effect, these measurements were made just before the $R_{sheet}(T)$ measurement for each $V_G$. The derived $n_{2D}$ and Hall mobility $\mu_{Hall}$ are summarized in Fig. 2C. The data in the $V_G$ range between 150 V and -40 V is telling. In this range, $n_{2D}$ only changes slightly whereas $\mu_{Hall}$ decreases from ~50 to ~15 cm$^2$V$^{-1}$s$^{-1}$. It strongly suggests that applying $V_G$ mainly modulates the disorder level in the conducting layer, which in turn modulates the electron scattering rate thereby the charge carrier mobility. The relatively fast decrease of $n_{2D}$ for $V_G < -40$ V is an indication of disorder-induced localization, and cannot be attributed to $\delta n_{2D}$ produced by capacitance effect (note that varying $V_G$ between -40 and -150 V can at best induce a $\delta n_{2D} \sim 0.6\times10^{13}$ cm$^{-2}$, which is only 37% of the experimentally observed value). It is convenient to characterize the disorder-level by the value of $k_F l$, where $k_F$ is the Fermi wave vector and $l$ is the electron mean-free path. With the help of the $R_{\text{sheet}}$(8 K) and $n_{2D}$(8.6 K) data, we found that $k_F l$ can be tuned from 17.5 ($k_F l \gg 1$, clean) to 0.4 ($k_F l \lesssim 1$, strongly dirty) by applying $V_G$ (from 150 V to -200 V, Fig. S9). Corresponding value of the Drude conductivity $\sigma_D^{(2D)} \sim (5-18) \times e^2/h$, at $k_F l > 5$.

As $V_G$ varies from 150 V to -80 V, the superconducting layer thickness decreases from above 6nm to below 2nm gradually (Fig. S10). As supported by this observation, the $V_G$-induced modulation of disorder level can be understood as sketched in Figs. 2D & 2E: the electron envelope wave function is pushed away or compressed towards the interface by a $V_G > 0$ or $< 0$. Because the disorder strength usually has a gradient from the dirty interface to the clean inside, a variation of the spatial distribution of electrons will equivalently cause a change in the average disorder level. It is worth mentioning that a similar modulation of the disorder level by applying $V_G$ was observed previously in LAO/SrTiO$_3$ interface (*19, 20, 22*). While the disorder effect generally plays a secondary role in LAO/SrTiO$_3$, it dominates in our LAO/KTO device. This discrepancy can be attributed to the following three factors: first, the dielectric constant of KTO(*23*) is smaller than that of STO(*1*); second, the $n_{2D}$ of LAO/KTO is larger than that of LAO/SrTiO$_3$(*1, 19–22*); third, compared with LAO/SrTiO$_3$(*19–22*), LAO/KTO has a much higher disorder level, as indicated by the low mobility $\mu_{\text{Hall}}$, probably due to the more disordered interface and the thinner superconducting layer.

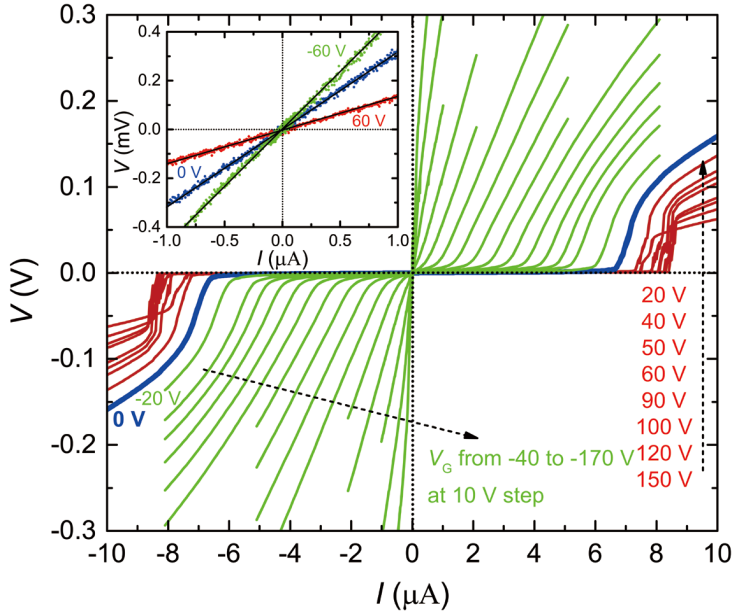

**Fig. 3. Current (*I*)-voltage (*V*) behaviors of the central 20 × 100 μm² bridge for $V_G$ varying from 150 to -170 V.** The measurements were performed at $T$ = 0.35 K. The evolution from superconducting to insulating states is indicated from the gradual variation of the superconducting critical current. Inset: Three typical linear *I-V* dependences in the low current range.

## Disorder-driven superconductor-to-insulator transition and the emergence of quantum metal

Taking above understanding, the extraordinary tuning achieved by $V_G$ is a disorder-driven effect in nature. Previously, disorder-driven superconductor-to-insulator quantum phase transition (QPT) has been intensely studied in disordered films (*24–27*). Compared with them, the present device is unique because its disorder level can be continuously tuned by $V_G$. Interestingly, although we have observed an apparent superconductor-insulator transition and a huge variation in $R_{sheet}$, a true superconducting ground state ($R_{sheet} = 0$) is absent. As shown in Fig. 2A (see also the Arrhenius plot of $R_{sheet}(T)$, Fig. S11), in the superconducting side, the resistance $R_{sheet}(T)$ does not reach zero even at lowest temperatures and saturates below ~1 K. Measured with or without filters, the residual resistance is unchanged, (Fig. S12), and can be modulated continuously by $V_G$, whose value spans in a surprisingly wide range from ~20 to above $10^3$ Ω/square.

The superconductivity in our system is justified not only by the significant drop of the resistance at $T_c$, but also by the existence of a critical current. As illustrated in Fig. 3, at $T = 0.35$ K, the critical current is ~8.3 μA at $V_G = 100$ V, and decreases as $V_G$ decreasing and vanishes eventually at the superconductor-insulator transition. For gating voltage $V_G$ from 150 to -130 V, the resistance at temperatures below $T_{BKT}$ (~1.7 K at $V_G = 0$) is measured by applying a DC current smaller than the critical current, where the linear I-V dependence (inset of Fig. 3) indicates an intrinsic ohmic resistance. As temperature is lowering, the resistance falls at first and then saturates at a value that can be orders of magnitude smaller than its normal state value and can be continuously tuned by varying $V_G$. Note that the resistance saturation begins at a relatively high temperature (~1 K), which together with its $V_G$-dependence and the linear I-V relation, exclude the possibility of the Joule heating. The residual resistance saturation in the superconducting side has been observed in diverse 2D superconducting systems (*8, 19, 24, 28–30*). The associated superconductors are called "failed superconductors" in literature (*30*). The corresponding ground state was called anomalous or quantum metallic state, whose origin is still under hot debates (*24, 27, 29–31*). The saturated resistance in the insulating sides may be ascribed to the applied-electric-field induced electron percolation on top of the insulating state. In this sense, the insulating state is fragile, and can be called "fragile insulator". It is worth mentioning that the resistance saturation in insulating states, although rare, was observed previously in disordered ultrathin Ga film (*32*), disorder-tuned LAO/SrTiO$_3$ interface (*19*), and graphene in a high magnetic field(*33*).

## $V_G$-modulation on Magnetic-field-driven QPT

In disordered 2D superconductors the magnetic-field-driven superconductor to insulator QPT (*24, 34*) is an interesting issue. The ability to continuously tune disorder level (from $k_F l \gg 1$ to $k_F l \sim 1$) in a single device enables us to monitor the disorder-dependent evolution of such QPTs. Figure 4A shows the plots of $R_{sheet}$ as a function of a perpendicular magnetic field $B$ for various temperatures below 2 K, at $V_G = 0$. These plots approximately cross at a single point, indicating a magnetic-field-driven superconductor to insulator QPT (*25, 26*). The crossing point gives rise to the critical field $B_c$ and the corresponding critical resistance $R_c$. To study the criticality of such QPTs, we analyzed the data with the scaling relation $R_{sheet}(\delta, T) = R_c f(\delta T^{-1/z\nu})$, where $\delta = |B-B_c|$, $f$ is an arbitrary function with $f(0) = 1$, $z$ is the dynamical critical exponent, and $\nu$ is the correlation length exponent (*7, 25, 27, 34, 35*). A striking observation is that the exponent product $z\nu$ is not universal, but goes divergently with decreasing temperature (Figs. 4B and its inset, S13, and S14). Similar divergent zv was observed previously in a few experimental studies(*35, 36*) and discussed by recent theoretical work (*30, 37*). One plausible explanation could be that the disorder in the conducting layer gives rise to superconducting granularities, on which local quantum phase fluctuations, rather than large-scale superconducting fluctuations, are relevant(*38*).

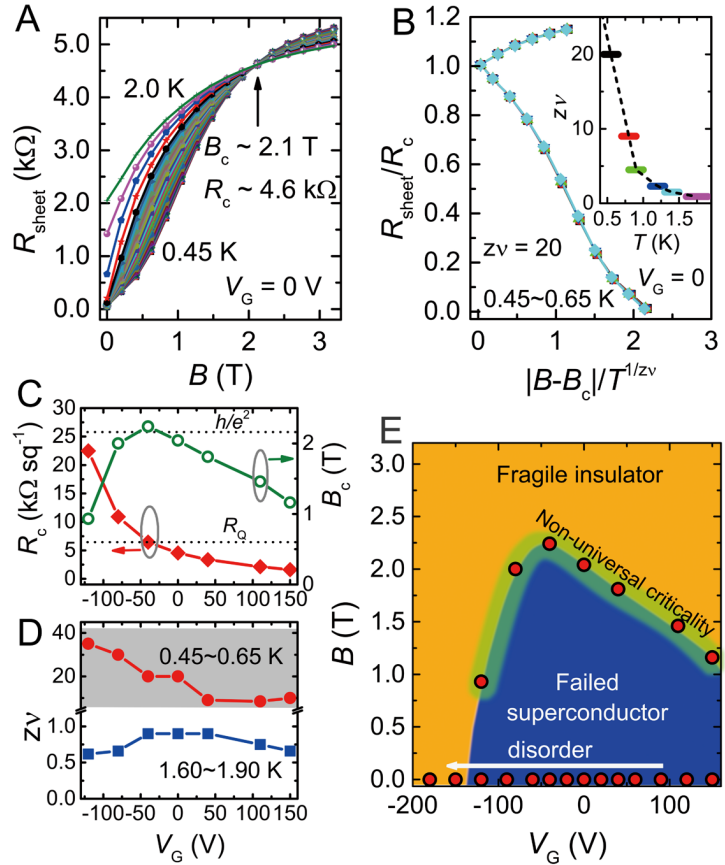

**Fig. 4. $V_G$-modulation on QPT driven by magnetic field perpendicular to the interface.** (**A**) Dependence of $R_{sheet}$ on magnetic field $B$ at fixed temperatures below 2 K. These curves were re-plotted from the temperature-dependent $R_{sheet}(T)$ data shown in Fig. S5A. (**B**) Scaling the data in 0.45 K ≤ $T$ ≤ 0.65 K with respect to a single variable $|B-B_c|/T^{1/zv}$, with $zv$ = 20. The inset shows the $zv$ values that are needed to scale the data in different temperature regimes (thick lines). (**C**) Dependences of $R_c$ and $B_c$ on $V_G$. The dashed lines indicate the positions of the quantum resistance $h/e^2$ and the pair quantum resistance $R_Q$. (**D**) Dependence of $zv$ on $V_G$ for 0.45 K ≤ $T$ ≤ 0.65 K (circles) and 1.60 K ≤ $T$ ≤ 1.90 K (squares). The grey shadow highlights the large uncertainty in determining the $zv$ values in the low temperature regime (see Fig. S14). (**E**) A schematic $V_G$-$B$ phase diagram for $T$ = 0. The red dots represent the experimental data generated from Figs. 2A & 4C. Note that for $B > B_c$, in the insulating phase, $R_{sheet}$ still saturates to a finite value (Fig. S15).

As summarized in Figs. 4C and 4D, by sweeping $V_G$ from 150 V to -120 V, the critical resistance $R_c$ is tuned continuously from ~1/4 to ~4 $R_Q$ (where $R_Q = h/4e^2 \approx 6.45$ kΩ is the quantum resistance of Cooper pairs), and the critical magnetic field $B_c$ exhibits a dome-shaped dependence on $V_G$, which reminds us of the dome-shaped $T_c$-$V_G$ relation in the absence of external magnetic field (Fig. 2B). All the $zv$ values in the high-temperature regime (1.60-1.90 K) are within 0.7 ± 0.2, which is similar to the ones observed in SrTiO$_3$-based 2D superconducting interfaces (*1*, *7*). In contrast, the $zv$ values in the low-temperature regime (0.45-0.65 K) are extremely large (divergent), implying that the non-universal criticality exists generally. Finally, based on all these observations, a schematic $V_G$-$B$ phase diagram is produced in Fig. 4E.

**Outlook**

The studies in the present work demonstrate that LAO/KTO(111) is an ideal platform to explore the rich physics of disordered 2D superconductors. New quantum electronic devices are expected to emerge with the unveiled unique electrostatic gating method.

**Acknowledgments:** We thank S. A. Kivelson, H. Y. Hwang, H. Yao, and F. C. Zhang for the fruitful discussions. **Funding:** This work was supported by the National Key Research and Development Program of China (grants 2017YFA0303002, 2016YFA0300202,

2016YFA0300204, 2016YFA0300701), National Natural Science Foundation of China (grants 11934016, 11774306, 11520101002, and 11921004), the Strategic Priority Research Program of Chinese Academy of Sciences (No. XDB28000000), and the Fundamental Research Funds for the Central Universities of China. **Author contributions:** Z.C., Y.L., H.Z., Y.S., and M.Z. prepared the samples; Z.C., Y.L., and Y.X. performed the transport measurements; Z.L. and H.T. performed the TEM measurements; Y.Z., J.S., and Y.X. analyzed the data and wrote the manuscript with input from all authors. **Competing interests:** The authors declare no competing interests. **Data and materials availability:** All data is available in the main text or the supplementary materials.

Supplementary Materials:

Materials and Methods

Figures S1-S16